\documentstyle[aps,preprint]{revtex}

\title{\Large{HIGHER-DERIVATIVE TWO-DIMENSIONAL MASSIVE FERMION THEORIES}}
\author{ \sc{ L. V. Belvedere$^\ast$,  R. L. P. G. Amaral$^\ast$,
C. G. Carvalhaes$^{\ast\ast}$ {\small and} N. A. Lemos$^\ast$} }
\address{
$\ast$ \small{ \it{Instituto de F\'\i sica, Universidade Federal
Fluminense\\ Av. Litor\^anea, s/n, Boa Viagem, Niter\'oi, CEP: 24210-340, RJ, Brasil.} }\\
$\ast \ast$\small{ \it{Instituto de Matem\'atica e Estat\'\i stica -
Universidade do Estado do Rio de Janeiro\\ Rua S\~ao Francisco Xavier,524, 
Maracan\~a, CEP: 20559-900, Rio de Janeiro, Brasil} }}
\date{\today}

\begin{document}


\maketitle

\begin{abstract}
We consider the canonical quantization 
of a generalized two-dimensional 
massive fermion theory containing higher odd-order derivatives. The 
requirements of Lorentz 
invariance, hermiticity of the Hamiltonian and 
absence of tachyon excitations 
suffice to fix the mass term, which contains a derivative coupling. We show 
that the basic quantum excitations of a higher-derivative theory 
of order $2N + 1$ consist of a physical usual massive fermion, quantized with 
positive metric, plus $2N$ unphysical massless fermions, quantized with 
opposite metrics. The positive metric Hilbert subspace, which is
isomorphic to the space of states of a massive free fermion theory,
is selected by a subsidiary-like condition.  Employing the 
standard bosonization scheme, the equivalent boson theory is derived. The 
results obtained are used as a guideline to discuss the solution of a theory 
including a current-current interaction.
\end{abstract}

\newpage

\section{Introduction }

There is a continuing interest in quantum field theories defined by higher-derivative Lagrangians \cite{Batlle88}. In spite of their possible shortcomings, such as ghost states and unitarity violation, field theories whose equations of motion are of order higher than the second are useful to regularize ultraviolet divergences \cite{Boulware84a}, especially for supersymmetric gauge theories \cite{Fayet77}. In certain higher-derivative string models, negative-norm states and unitarity violation can be avoided because it is possible to define a positive energy operator \cite{Nesterenko95}. The appearance of curvature-squared terms as corrections to the Einstein-Hilbert Lagrangian in the effective action of superstring theories \cite{Boulware84b} is a further reason why higher-derivative field theories are worth investigating for their own sake, and as such they have been studied from several different points of view in the last few years.  

Higher-derivative two-dimensional quantum field theories constitute a useful testing ground for powerful nonperturbative methods such as, for instance,
Fujikawa's technique in the context of path-integral quantization, or
exact operator solutions by means of bosonization in the framework of
standard canonical quantization \cite{Amaral93}. Recently, a 
higher-derivative generalization of the two-dimensional free fermion 
theory \cite{Amaral93,Belvedere95} has been constructed and exactly solved
by expressing the fermion fields of the model in terms of boson 
fields (``bosonization''). It turns out that the fermion fields that solve 
the higher-order equations of motion can be written in terms of usual 
Dirac fields, the so-called ``infrafermions''. Some of these 
infrafermions, however, need to be quantized with a negative metric, giving 
rise to an indefinite-metric Hilbert space.

In this paper we study the effect of the inclusion of a mass term on the 
behavior of these generalized fermion theories. We find that the requirements 
of Lorentz invariance, absence of tachyon excitations and hermiticity fix 
the form of the mass term, which differs from the usual one by the 
appearance of derivative couplings. The model is solved exactly and it so 
happens that the higher-derivative fermion fields admit only nonlocal 
mappings from usual fermion fields. With the help of the standard 
bosonization technique \cite{Mandelstam75,Halpern75}, the solution is 
expressed in terms of a sine-Gordon field and of two massless free scalar 
fields. These results are then employed to solve a theory with a 
current-current interaction.

The paper is organized as follows. In Section II we
introduce the third order massive fermion quantum field theory. In Section III 
we discuss it by functional methods. Starting with a so-called 
interpolating partition function \cite{Fradkin94}, and performing 
successive field transformations, the interpolating partition function is 
rewritten in terms of independent fields that satisfy equations of motion of 
order lower than the third. In this way we extract information regarding the 
physical content of the theory described  by the original third-order 
partition function. Guided by these results, in Section IV the third-order 
massive fermion theory is quantized by the canonical procedure. The physical Hilbert
space of positive defined norm,  which is isomorphic to the Hilbert
space of the free massive Fermi theory, is construct and the physical
states are accomodated as equivalent classes.  Section V is devoted to 
finding the equivalent boson theory. In Section VI a theory including a 
current-current interaction is discussed. Section VII is dedicated to 
general comments and conclusions.

\section{Description of the Model}

In a previous work \cite{Amaral93} we introduced a higher-derivative theory of a free massless fermion by means of the Lagrangian density \footnote{Our conventions are:
$$ \psi = ( \psi _{_{\!\!\ell}}, \, \psi _{_{\!\!r}} )^{\,T}, \,\, \epsilon^{\,0\,1} = g^{\,0\,0} = -g^{\,1\,1} = 1, \,\, \tilde \partial^{\,\mu } \equiv \epsilon^{\,\mu \,\nu} \partial_{\nu},\,\,\,\,\,\, x^\pm = \frac{x^0 \pm x^1}{2}, \,\,\, \partial_{_\pm} = \partial_{_0} \pm \partial_{_1}, $$
$$\gamma^{\,\mu} \gamma^{\,5} = \epsilon^{\,\mu \,\nu} \,\gamma _{\nu } ,\,\,\,
\gamma^{\,0} = \pmatrix{0&1 \cr 1&0}, \,\, \gamma^{\,1} = \pmatrix{0&1 \cr -1&0}, \,\, \gamma^{\,5} = \pmatrix{1&0 \cr 0&{-1}}\,\,. $$
}
\begin{equation}
{\cal L}^{(0)}(x) = -i \overline{\xi}(x) (\partial\!\!\!/ \partial\!\!\!/^{\dagger})^N   \partial\!\!\!/ \xi(x). 
\label{lag1}
\end{equation}
This model was analyzed and it was found that its basic excitations 
consist of $(2N + 1)$ independent canonical Dirac 
fields ({\it the infrafermions}), some quantized with positive 
metric and some with negative metric. These infrafermions were 
expressed in terms of scalar fields through the
Mandelstam formula, so that the bosonization mechanism was 
generalized for higher derivative 
fields. As an application, the gauged 
version (higher-derivative $QED_2$) was exactly solved. In usual first-order 
fermionic theories the
mass term appears as a nontrivial interaction term after bosonization. 
It is highly desirable to uncover the new features that arise from the introduction of a mass term for the higher-derivative fermion field. To this end we shall consider the Lagrangian density
\begin{equation}
{\cal L}_0(x) = -i \overline{\xi}(x) (\partial \!\!\!/ \partial \!\!\! / ^{\dagger}) ^N \partial \!\!\! / \xi(x) + m \overline{\xi}(x) (\partial \!\!\! / \partial \!\!\! /)^N \xi(x), \,\,\,\,\,\,\,\,\,\, (N>0),
\label{lag2}
\end{equation}
where $m$ is a parameter with dimension of mass.

In this Lagrangian density, the mass term is an even-order derivative coupling which prevents the appearance of tachyon excitations, preserves Lorentz invariance of ${\cal L}_0$ and provides a Hermitian Hamiltonian. Indeed, the Lagrangian density (\ref{lag2}) is a generalization of the usual massive Lagrangian and is obtained from it through the transformation 
\begin{equation}
\psi \rightarrow \gamma^{0} \partial \!\!\!/ \xi,
\label{xipsi}
\end{equation}
$\psi$ being the canonical massive free Dirac field. However, we remark that 
the
transformation (\ref{xipsi}) 
cannot be seen as a solution to the model mapping the higher 
derivative field $\xi$ to a lower
derivative (infrafermion) field $\psi$. It will 
be shown that the general quantum solution of the higher-derivative theory 
consist of a usual massive fermion, quantized with positive 
metric, plus $2N$ usual massless fermions, quantized with opposite 
metrics. We will find the expression for the field $\xi$ in terms of the 
infrafermions with canonical and
functional integral methods.  It means 
that (\ref{xipsi}) does not provide, upon inversion, a direct 
solution for the higher-derivative theory, since it does not allow the 
identification of the massless modes that contribute to the 
correlation functions of the phase space variables that define the
quantum theory and that generate the complete space of states.

The complexity involved here is greater than in the 
massless case \cite{Amaral93}, where the spinor 
components decouple and are treated independently. Therefore, in order to 
avoid unnecessarily complicated expressions, instead of 
considering the fairly general form (\ref{lag1}), we shall restrict 
ourselves to the third-order theory ($N=1$). In this case, one can 
verify, by solving the equations of motion, that a 
generalization of (\ref{lag1}) with mass term 
like $m \overline{\xi} \xi$ leads to tachyons, and no first-derivative mass 
term like $m \overline{\xi} \partial \!\!\!/ \xi$ respects Lorentz 
invariance, excluding non-local terms 
like $m \overline{\xi} \sqrt{\partial \!\!\!/ \partial \!\!\!/} \xi$.

Using the light-cone variables, the Lagrangian density is written as
\begin{equation}
{\cal L}_{_0}(x) = -i \xi_{_{(1)}}^*(x) \partial_{_-}^3 \xi_{_{(1)}}(x) - i 
\xi_{_{(2)}}^*(x) \partial_{_+}^3 \xi_{_{(2)}}(x) + m \xi_{_{(1)}}^*(x) \Box 
\xi_{_{(2)}}(x) + m \xi_{_{(2)}}^*(x) \Box \xi_{_{(1)}}(x).
\label{lag3}
\end{equation}
Following the discussion in \cite{Belvedere95}, under a Lorentz 
transformation $x^+ \rightarrow \lambda x^+$ and $x^- \rightarrow 
\lambda^{^{-1}} x^-$ we 
define $\xi_{_{(1,2)}} \rightarrow \lambda^{^{\mp 3/2}} \xi_{_{(1,2)}}$, so 
as to ensure Lorentz invariance of the theory.

\section{Functional Approach}

In order to obtain some insight on the general solution for the higher-derivative 
field $\xi$ in terms of infrafermions, we carry out a functional treatment. Our
starting point is the partition function
\begin{equation}
{\cal Z}_{_0} = \int D\overline{\xi} D\xi 
\exp \left\{ i \int d ^2 x \left[ -i \overline{\xi}(x) 
\partial \!\!\!/ \partial \!\!\!/ ^\dagger 
\partial \!\!\!/ \xi(x) + m \overline{\xi}(x) 
\partial \!\!\!/ \partial \!\!\!/ \xi(x) \right] \right\}.
\label{z0}
\end{equation}
In order to display a systematic functional procedure to decouple the 
higher-derivative theory by lowering the derivative order, we follow the
procedure of Ref.\cite{Fradkin94} and introduce an ``interpolating''
partition function ${\cal Z}_{_I}$, defined by
\begin{equation}
{\cal Z}_{_I} = \int D\overline{\xi} D\xi D\overline{\chi} D\chi 
\exp \left\{ i \int d ^2 x \left[ -i \overline{\chi} \partial\!\!\!/ \chi +  i \overline{\chi} \partial\!\!\!/ \gamma^0 \partial\!\!\!/ \xi - i \overline{\xi} \partial\!\!\!/ \gamma^0 \partial\!\!\!/ \chi +  m \overline{\xi} \partial\!\!\!/ \partial\!\!\!/ \xi \right] \right\}.
\label{zi}
\end{equation}
The partition function ${\cal Z}_{_I}$ describes an enlarged theory and is
connected to ${\cal Z}_{_0}$ by a weight factor corresponding to the partition
function of an extra ghost degree of freedom. This can be seen by performing in 
Eq.(\ref{zi}) the 
shift 
\begin{equation}
\chi = \chi^\prime + \gamma^0 \partial \!\!\!/ \xi\,,
\end{equation}
which yields
\begin{equation}
{\cal Z}_{_I} = {\cal Z}_{_0} \times \Big ( 
\int D\overline{\chi}^\prime D\chi^\prime e^{i \int d ^2x (-i 
\overline{\chi}^\prime \partial \!\!\!/ \chi^\prime) } \Big )\,.
\end{equation}
The additional degree of freedom is associated with a free massless 
fermion field quantized with negative metric. 

After having displayed the connection between the interpolating partition 
function ${\cal Z}_{_I}$ and the partition 
function ${\cal Z}_{_0}$, describing 
the original higher-derivative theory, we can 
use ${\cal Z}_{_I}$ as a starting point to
decouple the third-order theory by the successive lowering of the 
derivative order. 

First we shall reduce the order of the third-order derivative 
term. To begin with, we diagonalize the partition 
function ${\cal Z}_{_I}$ introducing
the transformation \footnote{This can be seen
as
$$
\xi = \xi^\prime + \frac{i}{m}\frac{1}{\Box}\partial\!\!\!/ 
\gamma^0 \partial \!\!\!/ \chi .
$$
In spite of nonlocality it is clear that there appears no 
Jacobian in this transformation.}
\begin{equation}
\partial \!\!\!/ \xi = \partial \!\!\!/ \xi^\prime + \frac{i}{m} \gamma^0 \partial \!\!\!/ \chi ,
\end{equation}
In this way we reduce ${\cal Z}_{_I}$, as given by (\ref{zi}), to a
partition function of a theory with a lowered derivative of second order,
\begin{equation}
{\cal Z}_{_I} = \int D\overline{\xi}^\prime D\xi^\prime D\overline{\chi} D\chi \exp \left\{ i \int d^2 x \left[ -i \overline{\chi} \partial \!\!\! / \chi - \frac{1}{m} \overline{\chi} \partial \!\!\! / \partial \!\!\! / \chi + m \overline{\xi}^\prime \partial \!\!\! / \partial \!\!\! / \xi^\prime \right] \right\}.
\label{zidesacoplado}
\end{equation}

The next step is the reduction of ${\cal Z}_{_I}$ to a first order 
theory. Following the same procedure, in order to obtain a further reduction 
of (\ref{zidesacoplado}), we introduce a second interpolating partition 
function ${{\cal Z}_{_I}}^\prime$, defined by
\begin{equation}
{\cal Z}_{_I}^\prime = \int D\overline{\xi}^\prime{D\xi^\prime}D\overline{\chi}{D\chi}D\overline{\psi}D\psi\exp\left\{i\int{d^2x}\left[-i\overline{\chi}\partial\!\!\!/\chi-m\overline{\psi}\psi-i\overline{\psi}\partial\!\!\!/\chi-i\overline{\chi}\partial\!\!\!/\psi+m\overline{\xi}^\prime\partial\!\!\!/\partial\!\!\!/\xi^\prime\right]\right\}.
\label{ziprime}
\end{equation}
The connection between the partion 
functions ${\cal Z}_{_I}^\prime$ and ${\cal Z}_{_I}$ can be displayed by 
performing in (\ref{ziprime}) the transformation 
\begin{equation}
\psi = \psi^\prime - (i/m) \partial \!\!\! / \chi\,,
\end{equation}
which yields
\begin{equation}
{\cal Z}_{_I}^\prime = {\cal Z}_{_I} \int D\overline{\psi}^\prime D\psi^\prime e^{ i \int 
d^2x (- m \overline{\psi}^\prime \psi^\prime)}.
\end{equation}
Since $\psi^\prime$ is not a dynamical field, the partition 
functions ${\cal Z}_{_I}^\prime$, given by (\ref{ziprime}), and ${\cal Z}_{_I}$, 
given by (\ref{zi}), have the same number of degrees of freedom.
 
Performing the transformation $\chi = \chi^{\prime\prime} - \psi$ in 
(\ref{ziprime}), we obtain
\begin{equation}
{\cal Z}^\prime_{_I} = \int D\overline{\chi}^{\prime\prime} 
D\chi^{\prime\prime} D\overline{\xi}^\prime D\xi^\prime 
D\overline{\psi} D\psi \exp \left\{ i 
\int d ^2x \left[ -i \overline{\chi}^{\prime\prime} \partial \!\!\! / 
\chi^{\prime\prime}  + i \overline{\psi} \partial \!\!\! / \psi 
- m \overline{\psi} \psi
+ m \overline{\xi}^\prime \partial \!\!\! / \partial \!\!\! / \xi^\prime \right] \right\}.
\label{ziprime2}
\end{equation}

In this way, we have reexpressed the interpolating partition 
function ${\cal Z}_{_I}$ in terms of a product of decoupled partition 
functions for the massless field $\chi^{\prime\prime}$, quantized with negative
metric, the canonical massive fermion field $\psi$ and a second order massless 
field $\xi^{\prime}$. We must remark that although the decoupling  
procedure to 
lowering the third-order term just replaces the original second-order mass 
term $ m \overline{\xi} \partial \!\!\! / \partial \!\!\! / \xi $ 
by $ m \overline{\xi}^\prime \partial \!\!\! / \partial \!\!\! / \xi^\prime $, its
presence in the original theory is crucial to perform the transition to the 
first-order fields $\chi^{\prime\prime}$ and $\psi$. 

The last step is the 
reduction of the second-order 
theory $ m \overline{\xi}^\prime 
\partial \!\!\! / \partial \!\!\! / \xi^\prime $ to first order. This can be 
done by adding a term $\delta \, \overline \xi ^\prime \partial
\!\!\!/ \xi ^\prime$ to 
the second order theory and, at the final stage of the derivative-order 
lowering procedure, taking the limit $\delta  \rightarrow 0$. This is
done in detail in Appendix B.

The counting of the resulting number of dynamical degrees of freedom 
involved in the interpolating theories gives two massless 
fields $ \chi^{\prime \prime} $, $ \xi^{\prime \prime} $, quantized with 
negative metric, one massless field $ \psi^\prime $, quantized with positive 
metric and one massive Dirac field $\psi$. Since the enlarged 
interpolating theories 
possess one more negative metric dynamical degree of freedom than the 
original 
theory, we can infer that the third-order theory described 
by ${\cal Z}_{_0}$ contains two massless fields quantized with 
opposite metrics and a massive free fermion field.

As a matter of fact, the usefulness of the decoupling procedure
depends, however, on generalizing it from the partition functions to the 
generating functionals. To this end we consider the generating functional
$$
{\cal Z}_{_0}[\overline{\eta}_i,\eta_i] = 
\int D\overline{\xi} D\xi \exp \left\{ i \int d ^2 x 
\left[ -i \overline{\xi}(x) \partial \!\!\!/ \partial \!\!\!/ ^\dagger 
\partial \!\!\!/ \xi(x) + m \overline{\xi}(x) 
\partial \!\!\!/ \partial \!\!\!/ \xi(x) +   
 \right. \right.
$$
\begin{equation}
\left. \left. + \overline{\eta}_1 \xi + \overline{\eta}_2 \,\gamma ^0
\partial \!\!\!/ \xi + \overline{\eta}_3\,\partial\!\!\!/^\dagger
\partial\!\!\!/ \xi + h. c.  \right] \right\}\,,
\end{equation}
This is the functional that generates the whole set of correlation functions 
of the phase space 
variables \footnote{\it See section III for 
definition of the phase space variables.} $\xi$, $ \gamma ^0 \partial \!\!\!/ \xi $ 
and $ \partial\!\!\!/^\dagger \partial\!\!\!/ \xi$. For the
interpolating theory, we consider the generating functional
$$
{\cal Z}_{_I} [\overline{\eta}_i,\eta_i,\overline{j},j] = \int 
D\overline{\xi} D\xi D\overline{\chi} D\chi 
\exp \left\{ i \int d ^2 x \left[ -i \overline{\chi} 
\partial\!\!\!/ \chi +  i \overline{\chi} \partial\!\!\!/ 
\gamma^0 \partial\!\!\!/ \xi - i \overline{\xi} \partial\!\!\!/ 
\gamma^0 \partial\!\!\!/ \chi +  \right. \right.$$
\begin{equation}
\left. \left. + m \overline{\xi} \partial\!\!\!/ \partial\!\!\!/ \xi +
\overline{\eta}_1 \xi + \overline{\eta}_2 \,\gamma ^0
\partial \!\!\!/ \xi + \overline{\eta}_3\,\partial\!\!\!/^\dagger
\partial\!\!\!/ \xi + h. c. + \overline{j} \chi + \overline{\chi} j \right] \right\}.
\end{equation}
Performing the transformation (7) on the interpolating generating
functional (16), although the source $j$  couples
with $\chi^\prime$ and $\gamma^0 \partial \!\!\!/ \xi$ we see that, up
to a normalization factor corresponding to the partition function of
the field $\chi^\prime$, the two 
generating functionals (15) and (16) are isomorphic in the space generated 
by the 
functional derivatives of the sources $\overline{\eta}_i$ and
$\eta_i$, leading to the same correlation functions. Thus, for any
polynomial $\mbox{\LARGE{\boldmath$\wp$}}$ of the functional derivatives
of $\eta_i$, we obtain
$$
\frac{1}{{\cal Z}_{_0}[0]}\,\mbox{\LARGE{\boldmath$\wp$}} 
\Big\{ \frac{\delta}{\delta \overline \eta_i},
\frac{\delta}{\delta \eta_i} \Big \} \,{\cal Z}_o \big [\overline
\eta_i,\eta_i \big ] \Bigg \vert_{\overline \eta_i = \eta_i = 0} =
$$
\begin{equation}
= \frac{1}{{\cal Z}_{_I}[0]} \,
\mbox{\LARGE{\boldmath$\wp$}} \Big\{ \frac{\delta}{\delta \overline \eta_i},
\frac{\delta}{\delta \eta_i} \Big \} \,{\cal Z}_{_I} \big [\overline
\eta_i,\eta_i,\overline j,j \big ] 
\Bigg \vert_{\overline \eta_i = \eta_i = \overline j = j = 0}\,.
\end{equation}

Following the same steps from (\ref{z0}) to (\ref{ziprime2}) we can
perform the decoupling mechanism for the generating functional. For
the source term associated with the field $\xi$ we get
\begin{equation}
\overline{\eta}_1 \xi = \overline{\eta}_1 \left[ \xi^\prime + \frac{i}{m} \frac{
\partial \!\!\!/ \gamma^0 \partial \!\!\!/}{\Box} (\chi^{\prime\prime} - \psi) \right].
\label{xixiprime}
\end{equation}
The correlation function for $\xi$ can be obtained from Eq.(\ref{xixiprime}). Indeed, making use of 
$\chi = \chi^{\prime\prime} - \psi$ we find 
\begin{eqnarray} 
\langle \xi \overline\xi \rangle_{_0} &=& \langle \xi^\prime
\overline\xi ^\prime  
\rangle_{_0} + \frac{1}{m^2} \langle \frac{\partial\!\!\!/ \gamma^0 
\partial\!\!\!/}{\Box} \chi \overline\chi \frac{\partial\!\!\!/ \gamma^0 
\partial\!\!\!/}{\Box} \rangle_{_0} \nonumber \\ 
&=& \frac{1}{m \Box} + \frac{1}{m^2} \frac{\partial\!\!\!/ \gamma^0 
\partial\!\!\!/}{\Box} \left( \frac{-1}{i\partial\!\!\!/ + \frac{1}{m} \Box} 
\right)  \frac{\partial\!\!\!/ \gamma^0 \partial\!\!\!/}{\Box} \equiv 
\frac{1}{-i  \partial\!\!\!/ {\partial\!\!\!/}^\dagger \partial\!\!\!/ + m 
\Box} 
\end{eqnarray} 
which coincides with the inverse of the operator that appears in 
the original formulation of the third order theory.

Note that Eq.(18) suggests that the representation of $\xi$ in terms of 
infrafermions  is nonlocal. This result will be confirmed by the 
operator approach to be  taken up next. We remark however that 
Eq.(\ref{xixiprime}) is not an infrafermion expansion for $\xi$. It provides 
the correct  dimensional and Lorentz properties of $\xi$ but its 
dynamical aspects are somewhat obscure  within this context. The physical 
aspects of the model shall become more clear within the canonical procedure.

\section{Canonical Quantization}

In order to perform the canonical quantization of the theory, we 
must obtain the basic Poisson brackets. In accordance with 
the third-order character of the Lagrangian 
density (\ref{lag3}), we 
take $\xi_{_{(1)}},\,\, \partial_{_-} \xi_{_{(1)}}, \,\,
\partial_{_-}^2 \xi_{_{(1)}} + i m \partial_{_+} \xi_{_{(2)}}, \,\, 
\xi_{_{(2)}}, \,\,\partial_{_+} \xi_{_{(2)}}, \,\, 
\partial_{_+}^2 \xi_{_{(2)}} + i m \partial_{_-} \xi_{_{(1)}}$ as the 
basic space phase variables. The associated canonical 
momenta \footnote{ Our choice for basic variables, different from the one 
in Ref.\cite{Amaral93}, has the advantage of providing momenta with 
homogenous Lorentz properties for the massive case.}, obtained by 
variation of the action around the equations of 
motion, are: $-i \partial_{_-}^2 \xi_{_{(1)}}^* - m \partial_{_+} \xi_{_{(2)}}^*, \,\, i \partial_{_-} \xi_{_{(1)}}^*, \,\, -i \xi_{_{(1)}}^*, \,\, -i \partial_{_+}^2 \xi_{_{(2)}}^* -  m \partial_{_-} \xi_{_{(1)}}^*,\,\, i \partial_{_+} \xi_{_{(2)}}^*, \,\, -i \xi_{_{(2)}}^*$, respectively. 

Using these variables, a systematic quantization (carried out using either a Dirac bracket formalism, if $\xi^{\dagger}$ is treated as an independent variable, or Poisson brackets, if $\xi^{\dagger}$ is taken as a function of $\xi$) furnishes the following nonvanishing equal-time anticommutators
\begin{eqnarray}
\nonumber &\{ \xi_{_{(1)}}(x) , \partial_{_-}^2 \xi_{_{(1)}}^*(y) \} = - 
\{ \partial_{_-} \xi_{_{(1)}}(x) , \partial_{_-} \xi_{_{(1)}}^*(y) \} = - \delta(x^1-y^1),&\\
&\{ \partial_{_-} \xi_{_{(1)}}(x) , \partial_{_+}^2 \xi_{_{(2)}}^*(y) \} = \frac{i}{m} \{ \partial_{_-}^2 \xi_{_{(1)}}(x) , \partial_{_-}^2 \xi_{_{(1)}}^*(y) \} = i m \delta(x^1-y^1).&
\label{anticom1}
\end{eqnarray}
The other nonvanishing equal-time anticommutators are obtained from (\ref{anticom1}) by switching the spinor indices and $x^{^1}$ to $- x^{^1}$.

Introducing the Fourier decomposition
\begin{equation}
\xi_{_{(\alpha)}}(x) = \int d^2k e^{-ikx} \tilde \xi_{_{(\alpha)}}(k),
\end{equation}
we obtain the general solution of the equations of motion
\begin{eqnarray}
\nonumber &\tilde \xi_{_{(1)}}(k) = a(k) \delta(k^2-m^2) + b_{_{(2)}}(k_{_-}) \delta(k_{_+}) + c_{_{(1)}}(k_{_+}) \delta(k_{_-}) - \frac{k_{_+}^2}{m} b_{_{(1)}}(k_{_+}) \frac{d}{dk_{_-}} \delta(k_{_-}),&\\
&\tilde\xi_{_{(2)}}(k) = \frac{k_{_-}^3}{m^3} a(k) \delta(k^2-m^2) + b_{_{(1)}}(k_{_+}) \delta(k_{_-}) + c_{_{(2)}}(k_{_-}) \delta(k_{_+}) - 
\frac{k_{_-}^2}{m} b_{_{(2)}}(k_{_-}) \frac{d}{dk_{_+}} \delta(k_{_+}).&
\label{sol1}
\end{eqnarray}

With the help of fields $\chi^i$, with dispersion relations described by
\begin{eqnarray}
\nonumber &\chi^1_{_{(1,2)}}(x) = -i \int d^2k \, a(k)\, k_{_{\mp}}\, \delta(k^2-m^2)\, e^{-ikx},&\\
\nonumber &\partial_{_\pm}^{^{-1}} \chi^2_{_{(1,2)}}(x) = \int dk_{_{\pm}}\, c_{_{(1,2)}} (k_{_{\pm}})\, e^{-ik_{_{\pm}} x^\pm},&\\
&\partial_{_\pm}^{^{-1}} \chi^3_{_{(1,2)}}(x) = \int dk_{_{\pm}}\,
b_{_{(1,2)}} \,(k_{_{\pm}})\, e^{-ik_{_{\pm}} x^\pm},&
\label{modea}
\end{eqnarray}
and consistently defining the operators \footnote{ A convenient definition for the non-local operator $\partial^{^{-1}}$ acting on the massless modes is $\partial_{_{\pm}}^{^{-1}} \chi_{_{(1,2)}} = 2^{-1} \int_{-\infty}^{x^\pm} dz \, \chi_{_{(1,2)}}.$ However, for calculating physical quantities, other definitions can be applied, so long as the identity $\partial_{_{\pm}} \left\{ \partial_{_{\pm}}^{^{-1}} \chi_{_{(1,2)}} \right\} = \chi_{_{(1,2)}}$
is ensured to hold.}
\begin{equation}
\partial_{_{\mp}}^{^{-1}}\chi^1_{_{(1,2)}}(x) = - \frac{1}{m^2} \partial_{_\pm} \chi^1_{_{(1,2)}}(x),
\end{equation}
we come back to the configuration space arriving at
\begin{eqnarray}
\nonumber
&\xi_{_{(1)}}(x) = \partial_{_-}^{^{-1}} \chi^1_{_{(1)}}(x) + 
\partial_{_+}^{^{-1}} \chi^2_{_{(1)}}(x^+) + \partial_{_-}^{^{-1}} 
\chi^3_{_{(2)}}(x^-) + i \frac{x^-}{m} \partial_{_+} \chi^3_{_{(1)}}(x^+),&\\
&\xi_{_{(2)}}(x) = \partial_{_+}^{^{-1}} \chi^1_{_{(2)}}(x) + 
\partial_{_-}^{^{-1}} \chi^2_{_{(2)}}(x^-) + \partial_{_+}^{^{-1}} 
\chi^3_{_{(1)}}(x^+) + i \frac{x^+}{m} \partial_{_-} \chi^3_{_{(2)}}(x^-).&
\label{sol2}
\end{eqnarray}

The mode $\chi^1$ is massive, whereas $\chi^2$ and $\chi^3$ are massless. In 
the general case (\ref{lag1}) this decomposition would generate one massive 
mode and $2N$ other massless modes. Tachyon excitations do not appear.

Inverting the relations (\ref{sol2}) and using the anticommutation 
laws (\ref{anticom1}) we obtain the following nonvanishing anticommutation relations
\begin{eqnarray}
\nonumber & \{ \chi^1_{_{(\alpha)}}(x) , {\chi^1_{_{(\alpha)}}}^{\!\!\! \dagger}(y) \} = \delta(x^1-y^1),& \\
\nonumber & \{ \chi^2_{_{(\alpha)}}(x) , {\chi^2_{_{(\alpha)}}}^{\!\!\! \dagger}(y) \} = - \frac{16}{m^4} \partial^4_{_1} \delta(x^1-y^1),& \\
 & \{ \chi^3_{_{(\alpha)}}(x) , {\chi^2_{_{(\alpha)}}}^{\!\!\! \dagger}(y) \} = - \frac{2 i \gamma^5_{\alpha\alpha}}{m} \frac{\partial}{\partial x^1} \delta(x^1-y^1).&
\label{anticom2}
\end{eqnarray}
The mode $\chi^1$ is a free massive Dirac field quantized with positive metric and the other two modes are noncanonical, $\chi^3$ being quantized with null metric. Nevertheless, the anticommutation structure (\ref{anticom2}) can be cast into a diagonal form by introducing a free massive field $\psi^1$ and two other free massless fields $\psi^2$ and $\psi^3$ quantized with opposite metrics
\begin{equation}
\{ \psi^1(x) , {\psi^1}^{\dagger}(y) \} = \{ \psi^2(x) , {\psi^2}^{\dagger}(y) \} = -\{ \psi^3(x) , {\psi^3}^{\dagger}(y) \} = 
\delta(x^1-y^1).
\end{equation}
In terms of these fields we have (see Appendix A),
\begin{eqnarray}
\nonumber
\chi^1_{_{(\alpha)}}(x) &=& \psi^1_{_{(\alpha)}}(x),\\
\nonumber
\chi^2_{_{(1,2)}}(x) &=& \left[ \frac{1}{2M^{^{p+1}}} \partial_{_{\pm}}^{^{p+1}} + (-1)^{^p} \frac{M^{^{p+1}}}{2m^4} \partial_{_{\pm}}^{^{3-p}} \right] \psi^2_{_{(1,2)}}(x)\\
\nonumber \mbox{} & & + \left[ \frac{1}{2M^{^{p+1}}} \partial_{_{\pm}}^{^{p+1}} - (-1)^{^p} \frac{M^{^{p+1}}}{2m^4} \partial_{_{\pm}}^{^{3-p}} \right] \psi^3_{_{(1,2)}}(x),\\
\chi^3_{_{(1,2)}}(x) &=& i(-1)^{^p} \frac{M^{^{p+1}}}{m} \partial_{_{\pm}}^{^{-p}} \psi^2_{_{(1,2)}}(x) - i(-1)^{^p} \frac{M^{^{p+1}}}{m} \partial_{_{\pm}}^{^{-p}} \psi^3_{_{(1,2)}}(x),
\label{modes3}
\end{eqnarray}
where $M$ is an arbitrary parameter of same dimension of $m$, and
$p$ an arbitrary integer. Under Lorentz transformations \cite{Belvedere95} 
we require that $M\rightarrow \lambda^{(1-p)/(p+1)}M$. Using this 
mapping, the original general solution turns out to be
\begin{eqnarray}
\nonumber \xi_{_{(1,2)}}(x) &=& - \frac{1}{m^2} \partial_{_{\pm}} 
\psi^1_{_{(1,2)}}(x) + i (-1)^{^p} \frac{M^{^{p+1}}}{m} 
\partial_{_{\mp}}^{^{-p-1}} \Big [ \psi^2_{_{(2,1)}}(x) - 
\psi^3_{_{(2,1)}}(x) \Big ] +\\
\nonumber &+& \left[ \frac{1}{2M^{p+1}} \partial_{_{\pm}}^{^p} + (-1)^{^p} \frac{M^{^{p+1}}}{2 m^{^4}} \partial_{_{\pm}}^{^{2-p}} - (-1)^{^p} \frac{M^{^{p+1}}}{m^{^2}} x^{\mp} \partial_{_{\pm}}^{^{1-p}} \right] \psi^2_{_{(1,2)}}(x)\\
&+& \left[ \frac{1}{2M^{p+1}} \partial_{_{\pm}}^{^p} - (-1)^{^p} \frac{M^{^{p+1}}}{2 m^{^4}} \partial_{_{\pm}}^{^{2-p}} + (-1)^{^p} \frac{M^{^{p+1}}}{m^{^2}} x^{\mp} \partial_{_{\pm}}^{^{1-p}} \right] \psi^3_{_{(1,2)}}(x)\,,
\label{sol3}
\end{eqnarray}
and is not a genuine operator-valued field. Note that it is impossible to adjust $p$ to describe the original fields 
locally in terms of usual fermions, while the corresponding relationship is 
local in the massless case \cite{Amaral93,Belvedere95}. However, the choice
$p=1$ appears to be particularly convenient since it
makes the parameter $M$ a scalar under Lorentz transformations.

The correlation functions do not depend on $p$ or $M$:
\begin{eqnarray}
\nonumber \langle 0 \vert \xi_{_{(1)}}(x) \xi_{_{(1)}}^{\ast}(y) \vert 0 \rangle &=& \frac{1}{m^4} \partial_+^x\partial_+^y S^{(+)}(x-y)_{_{11}} - \frac{i}{\pi m^4} \frac{1}{(x^+ - y^+ - i\epsilon)^3}\\ &-& \frac{i}{2 \pi m^2} \frac{x^- - y^-}{(x^+ - y^+ - i\epsilon)^2}\,\, ,\\
\langle 0 \vert \xi_{_{(1)}}(x) \xi_{_{(2)}}^{\ast}(y) \vert 0 \rangle &=& \frac{1}{m^4}\partial_+^x\partial_-^y  S^{(+)}(x-y)_{_{12}} + \frac{i}{4 \pi m} \int_{-\infty}^{x^+} dz^+ \frac{1}{z^+ - y^+ - i\epsilon}\,\, .
\end{eqnarray}
Wightman functions satisfying 
positive-definiteness and the cluster decomposition property can be 
obtained by  considering correlations between
appropriate derivatives of $\xi$, which represent genuine
operator-valued fields. For example:
\begin{eqnarray}
\nonumber \langle 0 \vert \partial_-\xi_{_{(1)}}(x)\partial_- \xi_{_{(1)}}^{\ast}(y) \vert 0 \rangle &=&  S^{(+)}(x-y)_{_{11}} \,\, ,\\
\langle 0 \vert \partial_-\xi_{_{(1)}}(x) \partial_+\xi_{_{(2)}}^{\ast}(y) \vert 0 \rangle &=&   S^{(+)}(x-y)_{_{12}}\,\, .
\end{eqnarray}
If one recalls the remarks concerning Eq.(3), it is not surprising that 
these derivatives of $\xi$ have the physical properties of the first order 
massive fermion field components. Indeed taking the corresponding
derivatives of Eqs.(\ref{sol3}) the massless infrafermions appear only in 
the combination $ (\psi^2-\psi^3) $ giving thereof no contribution
to the correlation functions. On the other hand, although  $\xi$  does not 
represent a genuine spinor operator-valued field the corresponding spinorial 
field nature is carried by the infrafermion 
fields $\psi^i$, which ensure for instance the correct microcausality  
requirements. In this sense, the general quantum field features of
the general solution $\xi$ are 
implemented by the infrafermion operators. The (infrafermion) massless 
operators are nevertheless nonlocal functions of the original fields. According 
to the general principles of QFT the Hilbert space of any theory
should be constructed from the basic field operators of the 
theory \cite{Morchio,Belvedereetal}. They make
up the basic building blocks, and in terms of {\it local} functions of them 
the algebra of fields  is defined. In a theory with 
fields obeying higher-derivative equations the algebra of fields should be 
enlarged by 
including all independent derivatives of the fields. Computing their 
correlation functions and then building
the Hilbert space is the route to discussing the physical content of the
theory. Under this perspective the infrafermion fields are in principle
artifacts to obtain the solution and it is not granted that they represent
``physical'' properties of the model. In our case the basic field is
$\xi$. But as we have seen $\xi$ is an unnaceptable field from wich to 
build a physical
Hilbert space since general rules of QFT such as positivity and cluster
decomposition property would be violated. To cure these pathologies 
let us construct the physical sub-algebra only through the derivatives
$\partial_-\xi_1$ and $\partial_+\xi_2$ and local functions of them. Since 
the massless 
fields do not contribute to correlation
functions of the derivatives of the field $\xi$ the algebra of 
fields becomes isomorphic to the algebra of the massive
infrafermion field. The higher-derivative theory is reduced {\it in
the free case} to the usual massive fermion field. 

The phase space variables and 
their associated momenta generate an indefinite metric 
Hilbert space ${\bf{\cal H}}$ violating the cluster decompositiom property. The 
{\it physical} Hilbert subspace ${\bf \widetilde {\cal H}}$ is selected by 
requiring that
$$ \widetilde{\cal H}\,\doteq\, \Big \{ \vert \Phi \rangle\,\in\,{\cal H}\,
\Big \vert \, ( \psi_{_{\alpha}}^2 - \psi_{_{\alpha}}^3 ) \vert \Phi \rangle 
= 0 \,\Big \} \,,$$
where the field combination $ (\psi^2 - \psi^3) $ generates from the 
vacuum $\mbox{\boldmath{$\Psi_o$}}$ zero norm states
$$ \langle \mbox{\boldmath{$\Psi_o$}} ( \psi_{_{\alpha}}^2(x) - \psi_{_{\alpha}}^3(x) )\,,\,
( \psi_{_{\alpha}}^2(y) - \psi_{_{\alpha}}^3(y) ) \mbox{\boldmath{$\Psi_o$}} \rangle = 0\,.$$
This condition ensures that the states generated from the 
derivatives of the field $\xi$
$$ (\partial^n\,\xi )\,\mbox{\boldmath{$\Psi_o$}} \,\in\,\widetilde{\cal H}\,\,,\,\,n \geq 1\,,$$
and the quocient 
space ${\bf\widehat{\cal H}} \doteq \widetilde{\cal H} / {\cal H}^o $ is 
isomorphic to the positive definite metric Hilbert space 
of the free massive Fermi theory.

\subsubsection{The Generator of Global Gauge Symmetry}

The conserved current associated with the global gauge symmetry is given by the products of fields and conjugate  momenta. The light-cone components  are
\begin{eqnarray}
\nonumber
&j^- = i \xi_{_{(1)}} (i \partial_{_-}^{^2} \xi_{_{(1)}}^* + m \partial_{_+} \xi_{_{(2)}}^*) + (\partial_{_-} \xi_{_{(1)}}) (\partial_{_-} \xi_{_{(1)}}^*) - i (-i \partial_{_-}^{^2} \xi_{_{(1)}} + m \partial_{_+} \xi_{_{(2)}}) \xi_{_{(1)}}^*,&\\
&j^+ = i\xi_{_{(2)}} (i \partial_{_+}^{^2} \xi_{_{(2)}}^* + m \partial_{-} \xi_{_{(1)}}^*) + (\partial_{_+} \xi_{_{(2)}}) (\partial_{_+} \xi_{_{(2)}}^*) - i (-i \partial_{_+}^{^2} \xi_{_{(2)}} + m \partial_{_-} \xi_{_{(1)}}) \xi_{_{(2)}}^*.&
\label{cur1}
\end{eqnarray}
Using these expressions and the diagonal expansions of $\xi$ and
dropping out total derivatives we obtain

\begin{equation}
j^\mu(x) = \overline{\psi}^1(x) \gamma^\mu \psi^1(x) + \overline{\psi}^2(x) 
\gamma^\mu \psi^2(x) - \overline{\psi}^3(x) \gamma^\mu \psi^3(x).
\label{cur2}
\end{equation}
 The charge operator becomes the sum of the charges of the infrafermions. Indeed defining
\begin{equation}
Q = \int{ dz^1 j^0(z) },
\label{charge}
\end{equation}
it is straightforward to show from (\ref{charge}), (\ref{sol3}) 
and (\ref{cur2}) that
\begin{equation}
\{Q,\xi(x)\} = -\xi(x).
\end{equation}

\section{Bosonization}

As emphasized in \cite{Belvedere95}, the Hamiltonian ${\cal H}_0$ obtained 
from the Legendre transformation of the Lagrangian (\ref{lag3}) evolves 
the $\xi$ field. The Hamiltonian ${\cal H}^\prime_0$ evolving the 
infrafermions is obtained from it by recognizing the time-dependent 
relationship between the basic variables and the infrafermions as a point 
transformation. The generating function may be constructed as 
in \cite{Belvedere95} and the Hamiltonian ${\cal H}^\prime_0$ computed. The 
result is the Hamiltonian for the three independent and canonical (except for metrics) first-derivative infrafermions:
\begin{equation}
{\cal H}^\prime_{_0} = -i \overline{\psi}^1 \gamma^1 \partial_{_1} \psi^1 - i \overline{\psi}^2 \gamma^1 \partial_{_1} \psi^2 + i \overline{\psi}^3 \gamma^1 \partial_{_1} \psi^3 + m \overline{\psi}^1 \psi^1.
\label{hhh}
\end{equation}
By means of a Legendre transformation one finds
\begin{equation}
{\cal L}^\prime(x) = \overline{\psi}^1(x) (i \partial \!\!\!/ -m ) \psi^1(x) + \overline{\psi}^2(x) (i \partial \!\!\!/) \psi^2(x) - \overline{\psi}^3(x) (i \partial \!\!\!/) \psi^3(x).
\label{lll}
\end{equation}
It is the Hamiltonian (\ref{hhh}) and the infrafermions that we are going to 
bosonize. The bosonization scheme we employ is the standard one  
\cite{Mandelstam75,Halpern75}. Using the Mandelstam representation 
\cite{Mandelstam75} for the Fermi field operators, we obtain

$$\psi^1_{_{(\alpha)}}(x) = (\frac{\mu}{2\pi})^{^{1/2}} e^{\,- i
\frac{\pi}{4} \gamma^5_{\alpha \alpha}}\,{\cal K}(\phi_2)\,
{\cal K}(\phi_3)\, :e^{-i \sqrt{\pi} 
\{\int_{-\infty}^x dz^1 \pi_{_1}(z) + \gamma^5_{_{\alpha\alpha}} \phi_{_1}(x)\}}:
\,,$$
\begin{equation}
\psi^2_{_{(\alpha)}}(x) = (\frac{\mu}{2\pi})^{^{1/2}} {\cal K}(\phi_1)\,
{\cal K}(\phi_3)\, :e^{-i \sqrt{\pi} 
\{\int_{-\infty}^x dz^1 \pi_{_2}(z) + \gamma^5_{_{\alpha\alpha}} \phi_{_2}(x)\}}:,
\label{bos1}
\end{equation}
$$\psi^3_{_{(\alpha)}}(x) = (\frac{\mu}{2\pi})^{^{1/2}} {\cal K}(\phi_1)\,
{\cal K}(\phi_2)\,\widehat{\cal K}(\phi_3)\, :e^{-i \sqrt{\pi} 
\{\int_{-\infty}^x dz^1 \pi_{_j}(z) + \gamma^5_{_{\alpha\alpha}} 
\phi_{_j}(x)\}}:,$$
where $\mu$ is an arbitrary finite mass scale, $\phi_{_2}$ and $\phi_{_3}$ are 
free and massless scalar fields, $\phi_{_1}$ is a sine-Gordon field 
and $\pi_{_j} = \dot\phi_{_j}$. The Klein factors that ensure 
the correct anticommutation relations between the independent fields
$\psi^j(x)$ are given by
\begin{equation}
{\cal K}(\phi_j) = e^{\,i \frac{\pi}{2}\,\int_{- \infty}^{+
\infty}\,d z^1\,\pi_j(z)}\,.
\end{equation}
Opposite metrics are also ensured by Klein factors \cite{Amaral93}. The Klein 
factor $\widehat{\cal K}(\phi_3)$ is introduced to ensure
the negative metric for the field $\psi^3(x)$ and is defined by
\begin{equation}
\widehat{\cal K}(\phi_3) = e^{\,i \pi\,\int_{- \infty}^{+
\infty}\,d z^1\,\pi_3(z)}\,.
\end{equation}

The equivalent boson field theory Hamiltonian is
\begin{equation}
{\cal H}^B_{_0} = \frac{1}{2} [\pi^2_{_1} + (\partial_{_1} \phi_{_1})^2] - 
\frac{m}{\pi} \mu \cos(2 \sqrt{\pi} \phi_{_1}) + \frac{1}{2} [\pi^2_{_2} + 
(\partial_{_1} \phi_{_2})^2] + \frac{1}{2} [\pi^2_{_3} + (\partial_{_1} \phi_{_3})^2].
\label{bbb}
\end{equation}
For the conserved current (\ref{cur2}) we find
\begin{equation}
\j^\mu(x) = -\frac{1}{\sqrt{\pi}} \varepsilon^{\mu\nu} \partial_{_\nu} 
\{\phi_{_1}(x) + \phi_{_2}(x) + \phi_{_3}(x)\}.
\end{equation}
The bosonization of the higher-derivative fermion fields $\xi (x)$ is obtained by 
using (\ref{bos1}) in (\ref{sol3}).

\section{Current-current Interaction}

The discussion in the final of section III suggests that the 
higher derivative theory, when implemented in such a way to avoid unphysical
properties, reduces to the first derivative theory. It is desirable to
address this issue when there is interaction. Under the functional integral
framework all decoupling procedure can be reproduced if the fermion
couples minimally with a gauge field. One should essentially replace usual
by covariant derivatives. The introduction of extra 
degrees of freedom in the decoupling partition function becomes now
more critical. We obtain the original partition function from the decoupling one discarding the partition function of the Schwinger
model with inverse fermion metric instead of the free fermion with
inverted metric. It is more suitable to proceed to a study in
 the canonical formalism.

In order to simplify matters let us consider here the current-current interaction. 
Consider the theory described by
\begin{equation}
{\cal L}_1(x) = {\cal L}_0(x) + g\,j^+(x)j^-(x),
\label{lc}
\end{equation}
where ${\cal L}_0$ is the Lagrangian density (\ref{lag3}), $j^{\pm}$ are 
given by (\ref{cur1}), $g$ is a constant and all the fields are in the 
Heisenberg picture. This is an extension of the Thirring model as a
third-order Lagrangian theory. A natural candidate to be the 
infrafermion Lagrangian density for this theory is built by adding the 
current-current interaction $j^\mu\,j_\mu$ with $j^\mu$ given by (20) to 
the Lagrangian density (\ref{lll}) in the Heisenberg picture. This identification is correct in the interaction picture, since the solution (\ref{sol3}) has led us to relate the third-order Lagrangian density (\ref{lag3}) to the first-order fermion theory (\ref{lll}) and the current (\ref{cur1}) to (\ref{cur2}). However, it is not clear that this direct identification remains true in the Heisenberg picture. It depends on generalizing the solution (\ref{sol3}), an issue we do not address here. In order to gain 
insight into this new theory we shall consider the first-order fermion theory
\begin{equation}
{\cal L} = \overline{\Psi}^1(i\partial\!\!\!/ - m)\Psi^1 + 
\overline{\Psi}^2(i\partial\!\!\!/)\Psi^2 -\overline{\Psi}^3 
(i\partial\!\!\!/)\Psi^3- g(\overline{\Psi}^1 \gamma^{\mu} \Psi^1 + 
\overline{\Psi}^2 \gamma^{\mu} \Psi^2 - \overline{\Psi}^3 
\gamma^{\mu} \Psi^3)^2.
\label{ifti}
\end{equation}
From now on, we shall use lowercase letters to denote fields in the 
interaction picture and the uppercase ones to those in the 
Heisenberg picture. We have thus been led to a Thirring model with 
global $SU(2,1)$ symmetry explicitly broken by the mass term.

Following \cite{Halpern75}, in the interaction picture the current-current 
term (\ref{cur2}) should be written as
\begin{equation}
{\cal H}^B_I = \frac{g}{2}\left[(j_F^0)^2 - \lambda(j_F^1)^2\right],
\end{equation}
where $\lambda$ is a parameter that has to be introduced in the 
interaction picture and is fixed by requiring Lorentz invariance. The 
subscript $F$ was inserted in order to emphasize that $j^\mu$ is a 
functional of free quantities, since we are in the interaction picture. After 
bosonization we find that
\begin{equation}
{\cal H}_I = \frac{g}{2} \left\{  (\partial_1\phi_1 + \partial_1\phi_2 - 
\partial_1\phi_3)^2 - \lambda (\pi_1 + \pi_2 - \pi_3)^2 \right\}.
\end{equation}
The full Heisenberg picture bosonized Hamiltonian density is immediately found:
\begin{eqnarray}
\nonumber {\cal H} &=& {\cal H}^\prime_0[\Phi,\Pi] + {\cal H}_I[\Phi,\Pi] = 
\frac{1}{2} \left[\Pi_1^2 + (\partial_1\Phi_1)^2\right] - 
\frac{m}{\pi}\,\mu\, \cos(2\sqrt{\pi}\Phi_1) + \frac{1}{2} \left[\Pi_2^2 + 
(\partial_1\Phi_2)^2\right]\\
&+& \frac{1}{2} \left[\Pi_3^2 + (\partial_1\Phi_3)^2\right] +
\frac{g}{2} (\partial_1\Phi_1 + \partial_1\Phi_2 - \partial_1\Phi_3)^2  - 
\frac{g\lambda}{2} (\Pi_1  + \Pi_2 - \Pi_3)^2.
\end{eqnarray}
The Heisenberg picture momenta $\Pi_i$ are
\begin{eqnarray}
\nonumber
\Pi_1 &=& \frac{1-2g\lambda}{1-3g\lambda}\dot\Phi_1 + \frac{g\lambda}
{1-3g\lambda}\dot\Phi_2 - \frac{g\lambda}{1-3g\lambda}\dot\Phi_3,\\
\nonumber
\Pi_2 &=& \frac{g\lambda}{1-3g\lambda}\dot\Phi_1 + \frac{1-2g\lambda}
{1-3g\lambda}\dot\Phi_2 - \frac{g\lambda}{1-3g\lambda}\dot\Phi_3,\\
\Pi_3 &=& - \frac{g\lambda}{1-3g\lambda}\dot\Phi_1 -
\frac{g\lambda}{1 - 3g\lambda}\dot\Phi_2 + \frac{1-2g\lambda}{1-3g\lambda}
\dot\Phi_3,
\label{PPP}
\end{eqnarray}
and the fields in the two pictures are related by $A_H = U^\dagger A_I U,\,\,
\dot U = -i {\cal H}_I U$.

A Legendre transformation yields the full Heisenberg picture 
Lagrangian $\cal L$. Requiring Lorentz invariance of $\cal L$ we 
obtain $\lambda = \frac{1}{1+3g}$. This result could also be achieved by 
imposing Schwinger's condition \cite{Halpern75}. Thus,
\begin{eqnarray}
\nonumber {\cal L} &=& \frac{1+g}{2}(\partial_\mu\Phi_1)^2 + 
\frac{1+g}{2}(\partial_\mu\Phi_2 )^2 + \frac{1+g}{2}(\partial_\mu\Phi_3)^2 +
 g(\partial_\mu\Phi_1)(\partial^\mu\Phi_2)\\
&-& g(\partial_\mu\Phi_1)(\partial^\mu\Phi_3)-
g(\partial_\mu\Phi_2)(\partial^\mu\Phi_3) + \frac{m}{\pi}\,\mu\,
\cos(2\sqrt{\pi}\Phi_1).
\label{lag4}
\end{eqnarray}

The transformations
\begin{eqnarray}
\nonumber
\Phi_1 &=& \frac{\sqrt{1+2g}}{\sqrt{1+3g}} \Phi_1^\prime,\\
\nonumber
\Phi_2 &=& \frac{-g}{\sqrt{(1+2g)(1+3g)}} \Phi_1^\prime + 
\frac{\sqrt{1+g}}{\sqrt{1+2g}} \Phi_2^\prime,\\
\Phi_3 &=& \frac{g}{\sqrt{(1+2g)(1+3g)}} \Phi_1^\prime + 
\frac{g}{\sqrt{(1+g)(1+2g)}} \Phi_2^\prime + \frac{1}{\sqrt{1+g}} \Phi_3^\prime
\label{diag2}
\end{eqnarray}
applied to $\cal L$ cast it in the diagonal form
\begin{equation}
{\cal L} = \frac{1}{2}(\partial_\mu\Phi_1^\prime)^2 + 
\frac{1}{2}(\partial_\mu\Phi_2^\prime)^2 + \frac{1}{2}(\partial_\mu
\Phi_3^\prime)^2 + \frac{m}{\pi}\mu \cos(2 a \sqrt{\pi} \Phi_1^\prime),
\label{lag5}
\end{equation}
where $ a = \big [ (1 + 2 g)/( 1 + 3 g) \big ]^{1/2} $. The dynamics of the interacting infrafermions is described in terms of two  free massless scalar fields plus a scalar field that satisfies a sine-Gordon equation modified by the appearence of the parameter $a$. Note that this factor in the argument of the cosine term
differs from the corresponding factor in the usual Thirring model. This suggests
that after introduction of interaction the physical Hilbert space is not isomorphic to 
that of the first-order theory, in contrast to the free case. 

Having obtained the canonical scalar fields, let us derive the bosonized 
expression of the infrafermions in the Heisenberg picture, which  amounts to 
writing all operators in (\ref{sol3}) as Heisenberg field operators. This 
means to apply the transformations (\ref{PPP}) and (\ref{diag2}) on
\begin{equation}
\Psi^j_{(\alpha)}(x) = (\frac{\mu}{2\pi})^{1/2} :e^{-i\sqrt{\pi} 
\{\int_{-\infty}^x dz^1 \Pi_j(z) + \gamma^5_{\alpha\alpha}\Phi_j(x)\}}:.
\label{bos2}
\end{equation}
The dynamics of the fields $\Phi^{\prime}_i$ is found from (\ref{lag5}). From 
(\ref{bos2}) and (\ref{lag5}) all expectation values of infrafermion fields can
be computed. It is worthwhile to remark that in the general case (\ref{lag1}) 
one would be led to a Thirring model with $SU(N+1,N)$ explicitly broken global 
symmetry.

\section{Conclusion}

We have discussed the generalization of the massive fermion theory by 
introducing higher derivatives. The requirements of Lorentz 
symmetry, hermiticity of the Hamiltonian, and absence of tachyon 
excitations suffice to fix the mass term. The mode expansion of the fermion 
fields has been explicitly made and it has been seen that one needs two 
massless first-order (infra) fermion fields and one massive free
fermion field to express the solution in usual form. In contrast to the massless case the relation 
between
the higher-derivative field and the infrafermions is non-local. A family 
of (equivalent) solutions has been constructed but all of them are non-local 
in some degree. This nonlocal expression of the infrafermions
in terms of the original field precludes the interpretation of their associated
particles as belonging to the spectrum of the theory, irrespective of the
 negative metric problem. The spectrum reduces to the free massive fermion when
local acceptable derivatives of the field are choosen for the definition of the
physical sub-algebra of fields of the model. An
 interesting point is that, in spite of the non-local 
relationship among the fields, the charge operator is obtained from
 the sum of the currents associated with each 
infrafermion including the negative sign expected for the negative metric 
infrafermion.

 As an example of application we have bosonized the model resulting from the 
current-current interaction expressed in terms of the infrafermions. The new 
infrafermion fields have been obtained,  allowing the computation of any 
number of correlation functions. The bosonized model was written in terms of 
one massive and two massless scalar fields. The effect of the interaction appeared 
through a change in the value of the mass and in the dependence of the 
infrafermions in all scalar fields introduced.

The generalization of the model by considering coupling with a gauge field, as 
in Ref.\cite{Amaral93}, is presently under investigation. Due to the 
presence of derivatives in the mass term this generalization is not a trivial 
rewriting of the treatment of the massive Schwinger model.

\centerline{\small{\bf Acknowledgments}}

{\small The authors express their thanks to Conselho Nacional de 
Desenvolvimento Cient\'{\i}fico e Tecnol\'ogico (CNPq) and to 
Coordena\c c\~ao de Aperfei\c coamento de Pessoal de N\'\i vel Superior 
(CAPES), Brazil, for partial financial support.}

\newpage

\appendix 
\setcounter{equation}{0}

\section*{A}

\hfill

For simplicity, we shall concentrate on the first component alone and shall derive the results for the case
$p = 1$. The generalizations can be obtained by following the same procedure.
Our first step consists in finding combinations
\ $\alpha$ and $\beta$ of
$\chi_{_{(1)}}^2$ and $\chi_{_{(1)}}^3$ such that
\begin{eqnarray}
\{ \alpha(x^+) , \alpha^\ast(y^+) \} & = & 0, \label{aaa1}\\
\{ \beta(x^+) , \beta^\ast(y^+) \} & = & 0,\\
\{ \alpha(x^+) , \beta^\ast(y^+) \} & = & \delta (x^1 - y^1). \label{aaa3}
\end{eqnarray}
From the anticommutation relations (\ref{anticom2}) we get
\begin{eqnarray}
\{ \partial^{-1}_+ \chi_{_(1)}^2(x^+) , \partial^{-1}_+ {\chi_{_(1)}^2}^{\!\!\!
\ast} (y^+) \} & = & \frac{4}{m^4} \partial_{_1}^{^2} \delta(x^1 - y^1),\\ 
\{ \partial_+ \chi_{_(1)}^3(x^+) , \partial^{-1}_+ {\chi_{_(1)}^2}^{\!\!\!
\ast}(y^+)  \} & = & -\frac{i}{m} \partial_{_1}^{^2} \delta(x^1 - y^1). 
\end{eqnarray}

Defining
\begin{eqnarray}
\alpha(x^+) & = & M \partial^{-1}_+ \chi_{_(1)}^{^2} (x^+) + b \partial_+
\chi_{_(1)}^{^3} (x^+),\\ 
\beta(x^+) & = & c \partial_+ \chi_{_(1)}^{^3} (x^+),
\end{eqnarray}
the relations (\ref{aaa1})-(\ref{aaa3}) are satisfied if one takes  %
\begin{equation}
b = i \frac{M}{2 m^3}, \,\,\,\,\,\,\,\,\,\,\,\, c = i \frac{m}{M}.
\end{equation}
Now all one needs is to adjust $M$ in order to obtain
$\psi_{_(1)}^{^2}$ and
$\psi_{_(1)}^{^3}$ from the combinations $\alpha + \beta$ and $\alpha - \beta$,
respectively. Inverting these relations we obtain
(\ref{modes3}) with $p = 1$. 

\newpage

\appendix

\setcounter{equation}{0}

\section*{B}

\hfill

In order to decouple the second-order theory for $\xi^\prime$, we
consider the following partition function
\begin{equation}
{\cal Z} = \int\,D \overline \xi ^\prime D \xi ^\prime \exp \Big \{ i
\int\,d ^2 z \big [ - i \delta \overline \xi ^\prime \partial \!\!\!/
\xi ^\prime + m \overline \xi ^\prime \partial \!\!\!/ \partial
\!\!\!/ \xi ^\prime \big ] \Big \}\,,
\end{equation}
and the corresponding interpolating theory
\begin{equation}
{\cal Z}_{_I} = \int\,D \overline \xi ^\prime D \xi ^\prime 
D \overline \psi ^\prime D \psi ^\prime
 \exp \Big \{ i
\int\,d ^2 z \big [ - i \delta \overline \xi ^\prime \partial \!\!\!/
\xi ^\prime + \frac{\delta^2}{m} \overline \psi ^\prime \psi ^\prime
- i \delta \overline \psi ^\prime \partial \!\!\!/  \xi ^\prime - i
\delta \overline \xi ^\prime \partial \!\!\!/ \psi ^\prime \big ] \Big \}\,.
\end{equation}
The connection between $ {\cal Z}_{_I}$ and ${\cal Z}$ can be seen by
performing in the interpolating partition function the change of variables
\begin{equation}
\psi ^\prime = \psi ^{\prime \prime} + i \frac{m}{\delta} \partial
\!\!\!/ \xi ^\prime \,,
\end{equation}
which leads to
\begin{equation}
{\cal Z}_{_I} = {\cal Z}\,\times\,\Big ( \int \,D \overline \psi ^{\prime
\prime} D \psi ^{\prime \prime}\,e^{\,i \int d ^2 z \big [\,
\frac{\delta^2}{m} \,\overline \psi ^{\prime \prime} \psi ^{\prime
\prime} \,\big ] }\,\Big )\,.
\end{equation}
and $\psi^{\prime\prime}$ is not a dynamical field. 

Introducing in the partition function ${\cal Z}_{_I}$, given by (B2), the 
decoupling transformation
\begin{equation}
\xi ^\prime = \xi ^{\prime \prime} - \psi ^{\prime}\,,
\end{equation}
we obtain
\begin{equation}
{\cal Z}_{_I} = \int\,D \overline \xi ^{\prime \prime} D \xi ^{\prime \prime}
D \overline \psi ^\prime D \psi ^\prime
 \exp \Big \{ i
\int\,d ^2 z \big [ - i \delta \overline \xi ^{\prime \prime} \partial \!\!\!/
\xi ^{\prime \prime} + i \delta \overline \psi ^\prime \partial \!\!\!/ \psi ^\prime
 + \frac{\delta^2}{m}\, \overline \psi ^\prime  \psi ^\prime \big ] \Big \}\,.
\end{equation}
The regulating parameter $ \delta $ makes  $\psi^\prime$ a massive field with 
mass $\delta/m$.

The limit $ \delta \rightarrow 0 $ is singular for the correlation functions 
of the 
fields $\xi^{\prime\prime}$ and $\psi^{\prime}$ individually. Nevertheless, this
limit is well-defined for the correlation functions of the 
second-order field  $ \xi^{\prime} \doteq
\xi^{\prime\prime}-\psi^{\prime} $, which is the field that 
contributes for the correlation functions of the original 
third order field $\xi$. From 
Eq. (A.6) we can obtain the two point function of the 
field  $\xi^{\prime}$, which yields
\begin{equation}
\langle \mbox{\boldmath{$\Psi_o$}} \xi^{\prime},
\xi^{\prime} \mbox{\boldmath{$\Psi_o$}} \rangle =
\langle \mbox{\boldmath{$\Psi_o$}}(\xi^{\prime\prime}-\psi^{\prime}),
(\xi^{\prime\prime}-\psi^\prime) \mbox{\boldmath{$\Psi_o$}} \rangle =
\frac{1}{m} \Big [ \slash\!\!\!\partial \big ( \slash \!\!\!\partial
- \frac{i\delta}{m} \big ) \Big ]^{-1}\,.
\end{equation}
The limit $\delta\rightarrow 0$ is well defined for general $n$-point
functions of the field $\xi^\prime$.

\end{document}